\documentclass[twocolumn,multicol,showpacs,preprintnumbers,amsmath,amssymb]{revtex4}
\usepackage[dvips]{graphicx}
\usepackage{color} 
\usepackage{bm}
\usepackage{rotating}
\begin{document}

{\large\begin{rotate}{0}{\color{red}Published in: Phys. Rev. B {\bf 74}, 205318 (2006).}\end{rotate}}
\title{Quantum size effects in a one-dimensional semimetal} 
\author{Shadyar Farhangfar} 
\email{shadyar@cc.jyu.fi}
\affiliation{Nanoscience Center and Department of Physics, University of Jyv{\"a}skyl{\"a},   
FIN-40014 Jyv{\"a}skyl{\"a}, Finland}
\date{\today}
\begin{abstract}
We study theoretically the quantum size effects in a one-dimensional semimetal  by a Boltzmann 
transport equation. We derive analytic expressions for the electrical conductivity, Hall coefficient, 
magnetoresistance, and the  thermoelectric power in a nanowire. The transport coefficients of 
semimetal oscillate as the size of the sample shrinks. Below a certain size  the semimetal evolves 
into a semiconductor. The semimetal-semiconductor  transition is discussed quantitatively. The  
results should make a theoretical ground for  better understanding of transport phenomena in 
low-dimensional semimetals. They can also provide useful information while studying 
low-dimensional semiconductors in general.  
\end{abstract}

\maketitle
\section{Introduction}
Quantum size effects (QSEs) arise if the magnitude of at least one dimension of the sample is 
comparable to the de Broglie wavelength of carriers in the material. Experimental study of such 
effects started by  an investigation of semimetallic bismuth thin films a long time ago \cite{orgin66}, 
followed by a theoretical model to explain the measured data soon after \cite{sando67}.  Since then 
the field has preserved its freshness to date, expanding its domain to include the study of size 
effects in conventional metals  \cite{govindaraj86},  semiconductors \cite{arora81},  
superconductors \cite{guo03} and, since recently, nanoclusters \cite{schmelzer02}, carbon 
nanotubes \cite{avouris99} and  fullerenes  \cite{tomanek04}.  Recent advances in nanofabrication 
techniques have made it possible to  extend  experimental investigation of QSEs in various types 
of semimetallic structures from two-dimensional (2D) thin films to one-dimensional (1D) nanowires 
\cite{heremans98}-\cite{rogacheva05}. On the other hand, despite the advances on an 
experimental front, there is not a  theoretical model for quantitative understanding of measured  
data in such  structures.  Here, we generalize the theoretical 2D model of Sandomirski{\u\i} 
\cite{sando67}  to study the confinement phenomena in a 1D regime. Emphasis will be on QSEs in 
semimetals and on the semimetal-semiconductor (SM-SC) transition. To illustrate the applicability 
of the model, size dependence of some of the transport coefficients  in bismuth nanowires will be 
addressed.     

\section{Semimetallic regime}
In a semimetal the conduction and valence bands overlap by a value $\Delta$ (see Fig. 1). At low 
temperatures, $k_{\rm B}T\ll \Delta$, as the sample size shrinks  this overlap decreases, finally 
reducing to the separation of the bands and the formation of an energy gap ${\cal E}_{\rm g}$. This 
is an immediate consequence of an  alteration in the energy density of states. Let us consider  a  
wire with dimensions $w$, $t$, and $L$ (width, thickness, and length, respectively). The 
single-particle wave functions are given by  \cite{cylindrical} 
\begin{equation}
{\Psi}_{ij}(k)=\frac{2}{\sqrt{w t L}}\sin(\frac{i\pi  x}{w})
\sin(\frac{j\pi  y}{t})\exp{\left({\bf i} kz\right)}.
\end{equation}
The corresponding energies for electrons are
\begin{equation}
{{\cal E}}_{ij}^{e}(k)={\cal E}_{ij}^{e}
+({\hbar^2 {k^2}}/{2m_{z}^{e}}),
\end{equation}
where we have used the shorthand notation
\begin{equation}
{\cal E}_{ij}^{e}\equiv\frac{\hbar^2\pi^{2}i^2}{2m_{x}^{e}w^2}+
\frac{\hbar^2\pi^{2}j^2} {2m_{y}^{e}t^2}\equiv{{\varepsilon}_x^e}{i^2}+{{\varepsilon}_y^e}{j^2}.
\end{equation}
Using the expression for electron density of states 
\begin{equation}
g^{e}({\cal E})=\frac{s}{2\pi \hbar}\frac{1}{wt}\sqrt{\frac{m_{z}^{e}}{2}}
\sum_{i,j}{\Theta\left({\cal E}-{\cal E}_{ij}^{e}\right)}({{\cal E}-{\cal E}_{ij}^{e}})^{-1/2},
\end{equation}
and the Fermi-Dirac distribution for electrons with chemical potential $\mu^e$, $f^{e}({\cal 
E})=[1+\exp({\cal E}- {\mu}^{e})/k_{\rm B}T]^{-1}$, through the Sommerfeld expansion, we get for 
the volume concentration of electrons at low temperatures 
\begin{equation}\label{elconc}
n({\cal E})=\frac{s}{2\pi\hbar}\frac{1}{wt}\sqrt{2{m_z^e}}{\displaystyle\sum}_{i,j}
({\mu}^{e}-{\cal E}_{ij}^{e})^{1/2}. 
\end{equation}
Above,  $\Theta$ is the Heaviside step function and $s$ is the spin  degeneracy.  The derivation of 
an analogous expression for the concentration of holes $p({\cal E})$ is straightforward. The 
electron and hole chemical potentials   $\mu^e$ and $\mu^h$, measured  from the bottom of a 
conduction band and the top of the valence band with respect to the Fermi level $\mu$,  can now 
be obtained from the charge neutrality condition $n({\cal E})=p({\cal E})$. Writing the Fermi energy 
as ${\mu^e}\equiv{\mu^e_x}+{\mu^e_y}$, we arrive at 
${\mu^e_j}/{\varepsilon^e_j}={\mu^h_j}/{\varepsilon^h_j}$. Here $ j=x, y$. Moreover, using the 
relation  $\Delta={\mu^e}+{\mu^h}$,  we get for the partial 
chemical potentials  
\begin{figure}\label{Fig.1}
\begin{center}
\includegraphics[width=82mm] {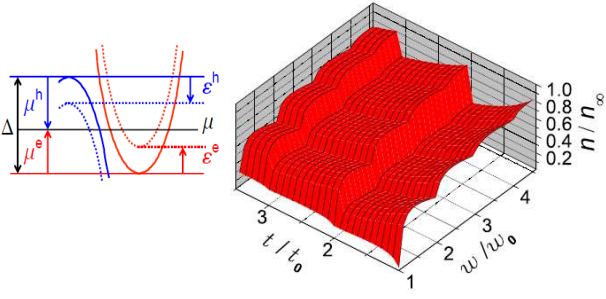}
\caption{Color online. {\bf left)} Overlapping bands in a semimetal. {\bf right)} Dependence of the  
carrier  concentration $n$, normalized by that  for a wire with ${w}/{w_0}=20$ and ${t}/{t_0}=20$, on 
the reduced width   and thickness. ${\Delta_x}=0.7\Delta$. The bismuth wire  is aligned along the 
bisectrix axes. The values for effective masses (in units of free-electron mass)  are    
${m_x^e}=0.00139$, ${m_y^e}=0.291$ and ${m_z^e}=0.0071$ for electrons, and 
${m_x^h}={m_y^h}=0.059$, and ${m_z^h}=0.634$ for  the holes.} 
\end{center}
\end{figure}
\begin{equation}
{\mu^e_j}={{\Delta}_j}{m_j^h}({{m_j^e}+{m_j^h}})^{-1}, \quad j=x, y.
\end{equation}
Above,  the energy overlap was represented as $\Delta\equiv{{\Delta}_x}+{{\Delta}_y}$.  
(Physically, ${\Delta}_x$ and ${\mu^e_x}$ can be interpreted as $x$ components of the bands 
overlap energy $\Delta$ and the Fermi level $\mu^e$ in the reciprocal lattice space; ${\Delta}_x$ is 
the overlap energy due to the confinement of carriers in the $x$ direction.) As the sample 
dimensions  become smaller, the conduction and valence bands slide upward and downwards 
respectively  and the band overlap $\Delta$  gradually diminishes.  The chemical potential $\mu$, 
however, remains  intact. This is illustrated in Fig. 1. Let  us assume that the band overlap vanishes 
at a width $w_0$ and at a thickness $t_0$, namely,  
${{\varepsilon}_x^e}({w_0})+{{\varepsilon}_x^h}({w_0})={{\Delta}_x}$ and 
${{\varepsilon}_y^e}({t_0})+{{\varepsilon}_y^h}({t_0})={{\Delta}_y}$.  Using these relations together 
with Eq. (6), we get for the width and thickness at which the semimetal-semiconductor transition 
happens 
\begin{equation}
{w_0}={\hbar\pi}/{\sqrt{2M_{x}{\Delta}_x}}, \quad{t_0}
={\hbar\pi}/{\sqrt{2M_{y}{\Delta}_y}}.
\end{equation} 
Here $\small{M_{j}}\equiv{m_j^e}{m_j^h}/({m_j^e}+{m_j^h})$ and $j=x, y$. The energy gap can  now  
be written as ${{\cal E}_{\rm g}}={{\Delta}_x}(\frac{w_0}{w})^2 + {{\Delta}_y}(\frac{t_0}{t})^2 -\Delta$. 
Designating $\small{{r_w}\equiv{w/{w_0}}}$ and $ \small{{r_t}\equiv{t/{t_0}}}$, with 
\begin{equation}
{{U}_{ij}^{e}}\equiv{1-({i}/{r_w})^2 +
({{\mu}_y^{e}}/ {{\mu}_x^{e}})\left[1-({j}/{r_t})^2\right]}, 
\end{equation}
we find for the normalized  electron concentration  
\begin{equation}
\frac{n}{n_{\infty}}=\frac{1}{wt}\sum_{m,n=1}^{[r_w],[r_t]}\sqrt{U_{mn}^e}
\left[\frac{1}{wt}{\displaystyle\sum}_{k,l=1}^{[r_w],[r_t]}{\sqrt{U_{kl}^e}}\right]_{w,t\rightarrow\infty}^{-1}
.
\end{equation}
Above, $[x]$ stands for the integer part of $x$. Figure 1  depicts the dependence of reduced 
electron concentration  on width and thickness.  Quantum size effects reveal themselves as steps in 
carrier concentration. By reducing the sample size the electron density reduces in steps ultimately 
becoming negligible below transition width (thickness). Within each step, however, the electron 
density varies nonmonotonously reaching its maximum value at a certain point.  The positions of 
these  maxima in each direction depend strongly on the effective masses, i.e., on the crystal 
structure of the wire. This is of vital importance from an experimental point of view; observation of 
the effects discussed below presumes a well-defined crystal orientation and due care must be paid 
to the fabrication of wires with comparable microscopic structures.

In what follows, we will utilize the Boltzmann transport equation (BTE) to derive various kinetic 
coefficients of interest  \cite{ziman}. Let  us mark the unperturbed and perturbed distribution 
functions with $f_0$ and $f$, the carrier charge with $q\in\{q_e, q_h\}$, and the velocity operator 
with $\hbar{\bf v}({\bf k})\equiv{\nabla}_{\bf k}{\cal E}({\bf k})$. In the presence of an external electric 
field ${\bf E}$, the BTE reads 
\begin{equation}
{\partial_t}f+{\bf v}\cdot{\nabla}_{\bf r}f+({q}/{\hbar} ){\bf E}\cdot{\nabla}_{\bf k}f=
\left({\partial_t} f\right)_{\rm scatt.}.
\end{equation}
Through the linearization of this equation and the introduction of a relaxation time $\tau$, the kinetic 
coefficients of a 1D system can now be obtained from 
\begin{equation}
({L}_{n})_{\alpha\delta}=\frac{s{q^2}}{2\pi}\int(-{{\partial}_{\varepsilon} f_{0}})
{\tau({\bf k})}({\cal E}({\bf k})-{\mu})^{n}{\bf v}_{\alpha}
{\bf v}_{\delta}{\rm d}{\bf k},
\end{equation}
where ${L_0}\equiv\sigma$ stands for the electrical conductivity and $S\equiv{L_1}/({qT}{L_0})$ is 
the thermoelectric power (Seebeck coefficient).  
\begin{figure}\label{Fig.2}
\begin{center}
\includegraphics[width=65mm] {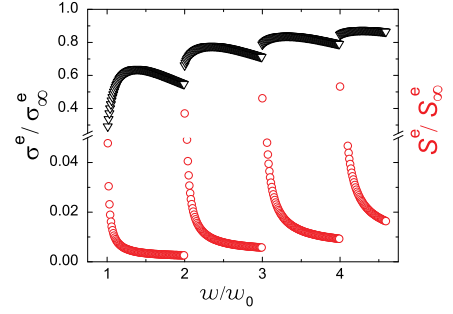}
\caption{Color online. Contribution of electrons to the electrical conductivity (triangles) and to the 
Seebeck coefficient (circles) of a bismuth wire versus its width at a fixed thickness $t/{t_0}=4.1$. 
(This corresponds to a thickness of about ${t_0}\approx 26$ nm). Other parameters are as given in 
Fig. 1.}
\end{center}
\end{figure}
To obtain an expression for the relaxation time we assume that $N$ scatterers, each with a 
strength $V_0$, are randomly distributed at positions ${\bf R}_j$ along the wire, 
$V({\bf r})={\sum}_{j=1}^{N}{V_0}\delta\left({\bf r}-{\bf R}_j\right)$,   and make use of the Fermi's 
golden rule 
${\tau}_{i\rightarrow{f}}^{-1}=({2\pi}/{\hbar}){\big|}
\langle {i\mid V({\bf r})\mid f}\rangle{\big|}^{2}g({\cal E}_f){\Omega}$. Here ${\Omega}\equiv{wtL}$ 
is the sample volume.  
Averaging over the configuration of scattering centers yields   
\begin{equation}
\frac{1}{{\tau}_{mn\rightarrow m'n'}^{e}}=\frac{\pi}{2}\frac{\varrho}{\hbar}{}
{V_0^2}{\Lambda}_{m'n'}^{mn}g^e({\cal E})\big| _{{\cal E}={\cal E}(k'm'n')},
\end{equation}
where  $\small{{\Lambda}_{m'n'}^{mn}\equiv(2+{\delta}_{mm'})(2+{\delta}_{nn'})}$  and 
$\varrho\equiv\frac{N}{\Omega}$ is the volume density of scatterers.  Electronic contribution to the 
electrical conductivity  now reads  
\begin{equation}
{\sigma}^{e}=\frac{2}{\pi}\frac{q_e^2}{\hbar}\frac{1}{\varrho} \frac{{\mu}_x^e}{m_z^e}
{\sum_{m,n=1}^{[r_w],[r_t]}}\frac{{\left({2\hbar}/{V_0}\right)^2}\sqrt{U_{mn}^e}}
{{\displaystyle\sum}_{m',n'=1}^{[r_w],[r_t]}{\Lambda}_{m'n'}^{mn}/\sqrt{U_{m'n'}^{e}}}. 
\end{equation}
Here the outer summation accounts for the contribution of different subbands $(m,n)$. Similarly, 
one obtains an analogous expression for the hole conductivity $\sigma^h$. The total electrical 
conductivity is the algebraic summation of electron and hole contributions. Figure 2 illustrates the 
contribution of electrons to the reduced electrical conductivity ${\sigma^e}/{{\sigma}_\infty^e}$ as a 
function of the wire width $w/w_0$.  
The Seebeck coefficient can be evaluated using the Cutler-Mott relation valid for a degenerate gas 
at low temperatures \cite{cutlermott69}, ${S_{\alpha\delta}}=\frac{\pi^2}{3}
\frac{k_{\rm B}^{2}T}{q}\frac{\rm d}{{\rm d}{\cal E}} {\ln\left[{g({\cal E})}v_{\alpha}({\cal 
E})v_{\delta}({\cal E}) \tau({\cal E})\right]}\big|_{{\cal E}=\mu}$.  Substituting for the corresponding 
quantities gives us    
\begin{equation}
{S^e_{mn}}=\frac{\pi^2}{6}\frac{{k_{\rm B}^2T}}{{q_e}}
\frac{1}{{\mu}_x^e}\left[\frac{1}{U_{mn}^e}+
\frac{\sum_{m',n'=1}^{[r_w],[r_t]}{\Lambda_{m'n'}^{mn}}(U^e_{mn})^{-3/2}}
{\sum_{m',n'=1}^{[r_w],[r_t]}{\Lambda_{m'n'}^{mn}}(U^e_{mn})^{-1/2}}\right].
\end{equation}
Dependence of the total electronic thermopower ${S^e}=\sum_{m,n}S_{mn}^{e}$ on the wire width 
is shown in 
Fig. 2. The  thermopower observable in the measurements, covering the contribution of both the 
electrons and the holes, is given by   $S=\frac{{\sigma^e} {S^e}+{\sigma^h} 
{S^h}}{{\sigma^e}+{\sigma^h}}$.      
\section{Semiconducting regime}
In the semiconducting regime the carrier gas becomes ultimately nondegenerate and the 
distribution function  will obey the Maxwell-Boltzmann (MB) statistics.  Denoting the distance of the  
Fermi level to  the bottom of the conduction band with $\mu^e$  and to the top of valence band with  
$\mu^h$ , the distribution function can now be expressed as $f^j({\cal E})\approx{\rm e}^{-\beta 
({\cal E}^j+\mu^j)}$ 
with $\beta\equiv{1/k_{\rm B}T}$ and $j=e,h$. Electron and hole concentrations for the intrinsic 
case, $n=p$ (per unit volume), are now obtained from 
\begin{eqnarray}
n(T)&=&\frac{s}{2}\frac{1}{{\hbar}{\sqrt\pi}}\frac{1}{wt}\left({{m_z^e}m_z^h}/{4}\right)^{1/4}
{\beta^{-1/2}}\\\nonumber
&&\times{\rm e}^{{-\beta{\cal E}_{\rm g}/2}}\left[{\displaystyle\sum}_{i,j;k,l}{\rm e}^{-\beta\left({\cal 
E}^e_{ij} + {\cal E}^h_{kl}\right)}\right]^{1/2}, 
\end{eqnarray}
with the gap energy given by ${\cal E}_{\rm g}={\mu^e}+{\mu^h}$. The chemical potential  $\mu$, 
through  the charge neutrality condition, reads 
\begin{eqnarray}
\mu(T)=\frac{{\cal E}_{\rm g}}{2}-\frac{1}{4\beta}\ln\small\left(\frac{m_z^e}{m_z^h}\small\right) 
-\frac{1}{2\beta}
\ln\small\left(\frac{{\sum}_{i,j}{\rm e}^{-\beta{\cal E}^e_{ij}}}{{\sum}_{k,l}{\rm e}^{-\beta{\cal 
E}^h_{kl}}}\small\right).
\end{eqnarray} 
Generally, the MB distribution makes the derivation of analytic expressions for the kinetic 
coefficients  more difficult and in most cases  one has to resort to numerical methods 
\cite{galvanomagnetic}. An alternative way, especially suitable at lower temperatures, would be to 
estimate  the quantities of interest with their average values.  For the  average value of the 
relaxation time  over energy distribution, $\left<{\tau}\right>\equiv
{{\int}\tau{\cal E}g({\cal E}){\partial_{\varepsilon} f}{\rm d}{\cal E}}\big/
{{\int}{\cal E}g({\cal E}){\partial_{\varepsilon} f}{\rm d}{\cal E}}$, assuming  
$({1}/{\left<{\tau}_{mn}\right>})={{\sum}_{m'n'}}({1}/{\left<{\tau}_{{mn}\rightarrow{m'n'}}\right>})$, we 
get 
\begin{eqnarray}
\left<\tau^e_{mn}\right>&=& \sqrt{\frac{2}{\pi}}\frac{1}{s}
\frac{wt}{\varrho}
\frac{\left({2{\hbar}}/{V_0}\right)^{2}}{{\sum}_{m',n'}{\Lambda}_{m'n'}^{mn}}
\frac{\beta^{-1/2}}{\sqrt{m^e_z}}\\\nonumber
&&\times\left[{{\displaystyle\sum}_{i,j}
\left(1+2{\beta}{\cal E}^e_{ij}\right){\rm e}^{-{\beta}{\cal E}^e_{ij}}}\right]^{-1}.
\end{eqnarray} 
Substituting for the relaxation time in Eq. (11), the electrical conductivity follows: 
\begin{eqnarray}
&{\sigma^e}=&
\frac{1}{2\pi}\frac{q_e^2}{\hbar}
\frac{1}{\varrho}\frac{1}{m_z^e\beta}\sum_{m,n}
\frac{\left({2\hbar}/{V_0}\right)^{2}}{{\sum}_{m',n'}{\Lambda}_{m'n'}^{mn}}
{\rm e}^{-{\beta}{\cal E}_{\rm g}/2}\\\nonumber
&&\times{{\rm e}^{-\beta\left({\mu^e}+{\cal E}^e_{ij}\right) }}
{\left[{\sum}_{i,j}
\left(1+2{\beta}{\cal E}_{ij}^{e}\right){\rm e}^{-{\beta}{\cal E}_{ij}^{e}}\right]^{-1} }.  
\end{eqnarray} 
The thermopower can be evaluated correspondingly, 
\begin{equation}
{S^e}=\frac{k_{\rm B}}{q_e}
\left[\frac{3}{2}+\beta{\mu^e}+
\frac{\sum_{m,n} {\beta}{\cal E}_{mn}^{e}{\rm e}^{-{\beta}{\cal E}_{mn}^{e}}}
{{\sum_{k,l} {\rm e}^{-{\beta}{\cal E}_{kl}^{e}}}}\right].
\end{equation} 
At low temperatures, in the absence of  intersubband scattering, one can limit himself to the 
contribution of the  first subband only, that is, ${\Lambda}_{m'n'}^{mn}={\Lambda}_{11}^{11}=9$,  
${\cal E}^e_{ij} \approx {\cal E}^e_{11}$ and ${\cal E}^h_{ij} \approx {\cal E}^h_{11}$. As a 
consequence, the electrical  conductivity has exponential dependence on the gap energy 
${\cal E}_{\rm g}$  (and thus on the wire dimensions) while the thermopower does not.  It would be 
in order to compare the expression above to that for a three-dimensional semiconductor with cubic 
symmetry given in Ref.\cite{ziman}, namely, ${S^e}=\frac{k_{\rm 
B}}{q_e}(\frac{5}{2}+c+\beta{\mu^e})$. Here, $c$ mirrors the energy dependence of  the relaxation 
time,  $\tau^{e}({\cal E})\propto{{\cal E}^c}$.   
\section{Galvananomagnetic  coefficients} 
Let  us assume that the wire is located in an external magnetic field ${\bf B}=(B_x,0,0)$ 
perpendicular to the wire axes and an external electric field $E_z$ is applied along the wire. The 
cyclotron frequency is marked with  ${\omega}_{\rm c}$.  Moreover, let the cyclotron radius be 
much larger than the lateral dimensions of the sample, $r_{\rm c}\gg{w, t}$. The Hall coefficient $R$ 
and the transverse magnetoresistance, 
$\frac{\Delta\rho}{{\rho}_0}\equiv\frac{\rho(B)-\rho(0)}{\rho(0)}$ with $\rho\equiv{1}/{\sigma}$,  can  
be easily  evaluated by employing the two-carrier model \cite{ziman,galvanomagnetic}. We obtain 
for the Hall coefficient 
\begin{equation}
R=\frac{{R^e}({\sigma^e})^2[1+({\sigma^h}{R^h})^2{B^2}] + 
{R^h}({\sigma^h})^2[1+({\sigma^e}{R^e})^2{B^2}]}
{({\sigma^e}+{\sigma^h})^2+{({\sigma^e}{\sigma^h})^2(R^e+R^h)^2{B^2}}}, 
\end{equation}
and for the magnetoresistance  
\begin{equation}
r\equiv\frac{\Delta\rho}{{\rho}_0}=
\frac{{\sigma^e}{\sigma^h}{(u^e+u^h)^2}B^2}{(\sigma^e+\sigma^h)^2+({\sigma^e}u^h-{\sigma^h}u^
e)^2{B^2}}.
\end{equation}
Above,   ${R^j}\equiv(n{q_j})^{-1}$ with $j=e,h$ is the single-particle Hall coefficient and $u^j$ 
stands for the carrier mobility.  Figure 3 illustrates the dependence of  the Hall coefficient and  
magnetoresistance  in weak magnetic fields  (${{\omega}_{\rm c}\tau}\ll{1}$) on the wire width at a 
fixed thickness.  Below, we compare some of the findings of our study  with the recent experimental 
data and with the other existing models. 
\begin{figure}\label{Fig.3}
\begin{center}
\includegraphics[width=65mm] {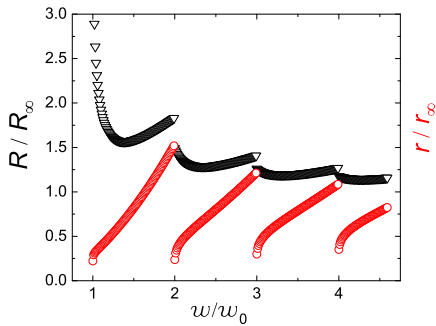}
\caption{Color online. Dependence of the  Hall coefficient (triangles)  and the magnetoresistance 
(circles) on the width in a weak magnetic field. Contributions from  both the electrons and the holes,  
as given by the two-carrier model, are taken into account.}  
\end{center}
\end{figure} 

\section{Discussion}
First, let us compare our model to a recent theoretical study  of  the SM-SC  transition in  cylindrical  
nanowires in which the electron and hole energies  are expressed  in terms of   the  Mathieu (or 
Bessel) functions \cite{bejenari04} . This  has to be compared to the expressions in Eq. (7).  While  
Ref.\cite{bejenari04} predicts the enhancement of QSEs as a consequence of anisotropy, our 
model foresees  sensitive variation of the SM-SC transition width (and thickness) and alteration in 
the form of carrier density and transport coefficients  as the values of effective masses change 
(Figs. 1-3). Furthermore, the model in Ref.\cite{bejenari04} misses the issue of transport. As it 
comes to practical applications, in most cases, the (lithographically fabricated) wires are of 
rectangular cross section as modeled in our study.  To make a quantitative comparison between 
predictions  of  our  model and the other existing ones, we consider a bismuth wire aligned along 
the bisectrix axis with effective masses given by  Ref.\cite{isaacson69} and presented in Fig. 1.  
The waveguide-analogous model in Ref.\cite{bejenari04} predicts a critical diameter of 
${d_c}\approx {32.5}$ nm for the SM-SC transition.  An earlier similar model, instead, gives 
${d_c}\approx{48.5}$ nm \cite{lin00}.  Correspondingly, in our model, substituting for the  values of 
effective masses   in Eq. (7) and with ${\Delta_x}=0.7\Delta$, we obtain  ${w_0}\approx{102}$ nm 
and ${t_0}\approx{26}$ nm.   Now, the diameter of an equivalent cylindrical wire, i.e. a wire with an  
equal cross section, can be  estimated as ${d_c}=\sqrt{{4w_0{t_0}}/{\pi}}\approx{58}$ nm.  

As is the case with the electrical conductivity and the Seebeck coefficient, at a fixed thickness, all 
the oscillations have the same periodicity given by Eq. (7). With  chosen parameters given above, 
this yields  ${w_0}\approx{102}$ nm. The  corresponding value observed in the experiments with a 
2D $n$-type bismuth film,  given in Ref.\cite{rogacheva03},  is  about ${100}$ nm. It is also a 
remarkable fact that, as predicted by the theory,  the measured value of film thickness for the 
semimetal-semiconductor transition equals the magnitude of periodicity near transition. However, 
in contrast to the predictions by our model,  in experiments  the period of oscillations becomes 
smaller as the transition thickness is approached. Also, the measured amplitudes of oscillations 
show less regularity than those  envisaged by the model. While part of the discrepancies  can be 
attributed to the structural nonidealities in the grown  films,  the others can be due to uncertainties in 
the chosen values for  the effective masses  and due  to the fact that the nondiagonal contributions 
in the effective mass tensor were ignored. Naturally, it would  be more desirable to compare the 
theoretical predictions to the experimental data in a single one-dimensional wire with a 
well-defined crystal orientation.  To our knowledge, despite recent experimental efforts with solitary 
nanowires of bismuth  \cite{chiu04,molares04},  the oscillatory dependence of kinetic coefficients 
on wire dimensions are yet to be observed. For the verification of predicted effects, we consider it 
crucial to have  nanowires  with comparable morphology,  and to characterize them with the 
precession four-probe measurements.

Dependence of the SM-SC transition on temperature and on the dopant concentration was 
investigated in a recent study of nanowires made of alloys of bismuth \cite{lin00}. There, a scheme  
based partly on the measurement of the Seebeck coefficient, was introduced to differentiate  
semimetallic nanowires from semiconducting ones. Correspondingly, in our model this can be 
performed by comparison of the form of the  measured thermopower to the forms of  Seebeck 
coefficients in SM and SC regimes, Eqs. (14) and (19),  respectively. At low temperatures the latter 
reduces to ${S^e}\approx\frac{k_{\rm B}}{q_e}(\frac{3}{2}+\beta{\mu^e}+\beta{\cal E}^{e}_{11})$ 
and depends inversely on temperature,  the former instead has a linear $T$ dependence. Also, at 
higher temperatures, the thermopower in semiconducting regime  saturates to  an asymptotic  
value, ${S^e}\approx\frac{3}{2}\frac{k_{\rm B}}{q_e}$,  comparable to (and slightly smaller than)  the 
bulk value ${S^e}\approx\frac{k_{\rm B}}{q_e}(\frac{5}{2}+c)$.  All these predictions agree with  the 
experimentally observed behavior of the thermopower  illustrated in Fig. 3 of  Ref.\cite{lin02}. Our 
model does not account for the effect of dopant concentration  on the transport properties of the  
alloys.  However, in principle, there is not any restriction to extend it further to  cover the doping 
effects too. 

Finally, let us briefly discuss the relevance of our study to the similar phenomena in  carbon 
nanotubes (CNTs).  Depending on its crystal structure, a CNT can be metallic or  semiconducting. 
Typically the energy gap of a semiconducting tube is small, $\sim 0.5$ eV.  Any disturbance  
(bending of the tube for example) can alter  the crystal structure and introduce a larger energy gap 
or reduce it; a metallic tube may turn semiconducting or vice versa. An analogy to the quantum size 
effects discussed above is apparent. Accordingly, we may anticipate the stepwise change of 
carbon nanotube conductance as  a function of mechanical disturbance. The implementation of 
such an idea could be of interest in  nanoelectromechanical applications; in designing  an 
ultrasensitive transducer,  for instance.   
\section{conclusions}
To summarize, a model for the study of confinement effects and transition to the semiconducting 
regime in one-dimensional semimetals is introduced. The model can be  extended to include the 
nonparabolic contributions to the dispersion relation and to cover doped semimetals and 
semiconductors at finite temperatures as well. Considering the nonmonotonic oscillatory 
dependence of the transport coefficients on the sample size, the analytic results presented here 
can be readily utilized to optimize various figures of merit in different applications.  Similar 
approaches to the one introduced here can be adopted for the study of size effects in other 
low-dimensional systems, such as quantum dots or mesoscopic  superconductors,  for example. 


\begin{thebibliography}{99}
\bibitem{cylindrical} For wires with a circular cross section, one can use a cylindrical potential well 
and obtain the dispersion relations in terms of the roots of Bessel functions \cite{lin00,bejenari04}. 
However, this will make the derivation of analytic expressions for the relaxation time (and kinetic 
coefficients) impossible. In this case, a reasonable approximation would be to substitute $w$ and 
$t$ of a rectangular wire with an equivalent diameter $\frac{{\pi}{d^2}}{4}=wt$.
\bibitem{galvanomagnetic}  Derivation of  analytic expressions for the transport coefficients in the  
nondegenerate regime, and/or in the presence of a magnetic field, is a tedious task and deserves 
a dedicated study.  In some cases, however,  one may  introduce a trial  solution   and derive all the 
components of a conductivity tensor in  a magnetic field.
\end{thebibliography}
\end{document}